\documentclass[12pt]{article}
\usepackage{graphics}
\begin{document}
\title{Quantum entanglement and Bell violation of two coupled cavity fields
in dissipative environment}
\author{Shang-Bin Li$^{a,b}$ and Jing-Bo Xu$^{a,b}$\\
$^{a}$Chinese Center of Advanced Science and Technology (World Labor-\\
atory), P.O.Box 8730, Beijing, People's Republic of China;\\
$^{b}$Zhejiang Institute of Modern Physics and Department of Physics,\\
Zhejiang University, Hangzhou 310027, People's Republic of
China\thanks{Mailing address}}
\date{}
\maketitle

\begin{abstract}
{\normalsize We study the quantum entanglement between two coupled
cavities, in which one is initially prepared in a mesoscopic
superposition state and the other is in the vacuum in dissipative
environment and show how the entanglement between two cavities can
arise in the dissipative environment. The dynamic behavior of the
nonlocality for the system is also investigated.\\

}
\end{abstract}

\newpage
Quantum entanglement has been recognized as an important resource
for quantum information processing [1,2]. Entanglement can exhibit
the nature of a nonlocal correlation between quantum systems that
have no classical interpretation. In recent years, the quantum
information processing by utilizing the cavity QED has attracted
much attention [3-7], and the entanglement generation and
nonlocality test of two cavity fields have been investigated
[8-11]. Meanwhile, the system of two distinct coupled cavities in
dissipative environment has been proposed to study the reversible
decoherence of mesoscopic superposition states [12,13]. On the
other hand, there have been theoretical and experimental studies
on quantum nonlocality for continuous variables, in terms of
Wigner representation in phase space based upon parity measurement
and displacement operation [14-16], or by using the pseudospin
Bell operator [17,18]. In this Letter, we study the quantum
entanglement between two coupled cavities, in which one is
initially prepared in a mesoscopic superposition state and the
other is in the vacuum in dissipative environment. We show how the
entanglement between two distinct cavities can be generated in the
dissipative environment. The influence of different initial states
of the cavities on the entanglement is also discussed.
Furthermore, we investigate the dynamic behavior of the quantum
nonlocality based on parity measurement for the two coupled cavity
fields in a dissipative environment.\\
\hspace*{8mm}We consider the situation that two distinct cavities
are coupled in the dissipative environment. The master equation
describing the time evolution of the cavity fields under the usual
Born-Markov approximation is given by [12,13],
$$
\frac{d\rho(t)}{dt}=-(i\omega+k)a^{\dagger}a\rho(t)+(i\omega-k)\rho(t)a^{\dagger}a
+2ka\rho(t)a^{\dagger}
$$
$$
~~~-(i\omega+k)b^{\dagger}b\rho(t)+(i\omega-k)\rho(t)b^{\dagger}b+2kb\rho(t)b^{\dagger}
$$
$$
~~~-i\gamma[a^{\dagger}b\rho(t)-\rho(t)a^{\dagger}b]-i\gamma[b^{\dagger}a\rho(t)
-\rho(t)b^{\dagger}a],
\eqno{(1)}
$$
where $k$ denotes the decay constant, $\gamma$ is the coupling
constant of two cavity fields, $a$ ($a^{\dagger}$) is the
annihilation (creation) operator of cavity field $A$, $b$
($b^{\dagger}$) is the annihilation (creation) operator of cavity
field $B$, and $\omega$ is the frequency of cavity fields referred
$A$ and $B$, two resonating identical cavities. In Ref.[12,13],
the cavity field $B$ is regards as a well-controlled single mode
environment. In some extents, the entanglement between cavity
fields $A$ and $B$ can be regarded as nonclassical correlation
between a cavity field and its well-controlled single mode
environment. It is easy to obtain the explicit analytical solution
of Eq.(1) as follows,
$$
\rho(t)=\sum^{\infty}_{n_1=0,n_2=0}\frac{[1-\exp(-2kt)]^{n_1+n_2}}{n_1!n_2!}U_1U_2a^{n_1}b^{n_2}\rho(0)a^{\dagger{n_1}}b^{\dagger{n_2}}U_2U^{\dagger}_1,
\eqno{(2)}
$$
where
$$
U_1=\exp[-it(\omega{a}^{\dagger}a+\omega{b}^{\dagger}b+\gamma{a}^{\dagger}b+\gamma{b}^{\dagger}a)],
\eqno{(3)}
$$
and
$$
U_2=\exp[-kt(a^{\dagger}a+b^{\dagger}b)]. \eqno{(4)}
$$
Firstly, we assume that the cavity field $A$ is
prepared in a single photon Fock state $|1\rangle$ and the cavity
field $B$ is prepared in the vacuum state. Then, the time
evolution density operator $\rho(t)$ can be written as
$$
\rho(t)=(1-e^{-2kt})|00\rangle\langle00|+e^{-2kt}|\psi\rangle\langle\psi|,
\eqno{(5)}
$$
where
$|\psi\rangle=\cos\gamma{t}|10\rangle-i\sin\gamma{t}|01\rangle$.
The concurrence [19] quantifying the entanglement between two
cavity fields described by $\rho(t)$ in Eq.(5) can be easily
obtained
$$
C=|\exp(-2kt)\sin2\gamma{t}|, \eqno{(6)}
$$
where $|x|$ gives the absolute value of $x$. One may expect that
there are some relations between the entanglement and mixedness of
cavity field $A$. In fact, a simple equation can be derived to
characterize the relation between the entanglement and mixedness
of cavity field $A$. If we quantify the mixedness of cavity field
$A$ by making use of the linear entropy defined by,
$$
M=1-{\mathrm{Tr}}\rho^2_A, \eqno{(7)}
$$
where $\rho_A$ is the reduced density matrix of the cavity field
$A$. For the density matrix $\rho(t)$ in Eq.(5), the linear
entropy $M$ is obtained
$$
M=2e^{-2kt}\cos^2\gamma{t}(1-e^{-2kt}\cos^2\gamma{t}). \eqno{(8)}
$$
We can see that the purity of the cavity field $A$ is directly
related with the entanglement between the cavities $A$ and $B$ in
the dissipation.\\

Then, we assume that the cavity field $A$ is initially prepared in
a pure state
$|\phi\rangle=\cos(\theta/2)|1\rangle+\sin(\theta/2)e^{i\tau}|0\rangle$
or a mixed state
$\cos^2(\theta/2)|1\rangle\langle1|+\sin^2(\theta/2)|0\rangle\langle0|$,
and the cavity field $B$ is still in the vacuum state. We can see
that the two kinds of initial conditions cause the same
entanglement evolution. We can find the time evolution density
matrix
$$
\rho(t)=[1-e^{-2kt}\cos^2(\theta/2)]|00\rangle\langle00|
+\cos^2(\theta/2)\exp(-2kt)|\psi\rangle\langle\psi|
$$
$$
~~~+\frac{1}{2}\sin\theta\exp(-kt)[\exp(-i\omega{t}-i\tau)|\psi\rangle\langle00|
+\exp(i\omega{t}+i\tau)|00\rangle\langle\psi|], \eqno{(9)}
$$
for pure initial state and
$$
\rho(t)=[1-e^{-2kt}\cos^2(\theta/2)]|00\rangle\langle00|
+\cos^2(\theta/2)\exp(-2kt)|\psi\rangle\langle\psi|, \eqno{(10)}
$$
for mixed initial state, respectively. The entanglement quantified
by concurrence of fields $A$ and $B$ for two kinds of initial
conditions have the same expression as follows,
$$
C=|\exp(-2kt)\cos^2(\theta/2)\sin(2\gamma{t})|. \eqno{(11)}
$$
From Eq.(11), we can see that the entanglement is proportional to
the particle population of cavity field $A$ in the single photon
Fock state $|1\rangle$.\\
\hspace*{8mm}Next, we discuss two kinds of entangled state of two
cavities in dissipative environment. We assume that the initial
state of two cavity fields is in one of four Bell states, i.e.,
$\rho(0)=|B^{i}\rangle\langle{B^{i}}|$, where the four Bell states
$|B^{i}\rangle$ ($i=1,2,3,4$) are defined by
$$
|B^{1}\rangle=\frac{\sqrt{2}}{2}(|11\rangle+|00\rangle),
$$
$$
|B^{2}\rangle=\frac{\sqrt{2}}{2}(|11\rangle-|00\rangle),
$$
$$
|B^{3}\rangle=\frac{\sqrt{2}}{2}(|10\rangle+|01\rangle),
$$
$$
|B^{4}\rangle=\frac{\sqrt{2}}{2}(|10\rangle-|01\rangle).
\eqno{(12)}
$$
It is easy to prove that $|B^{1}\rangle$ and $|B^{2}\rangle$ in
the cavities are more fragile against dissipative environment than
$|B^{3}\rangle$ and $|B^{4}\rangle$. The four expressions of
concurrence of the time evolution density matrix corresponding
four different initial Bell states can be written as
$$
C^{(1)}=\exp(-4kt),
$$
$$
C^{(2)}=\exp(-4kt),
$$
$$
C^{(3)}=\exp(-2kt),
$$
$$
C^{(4)}=\exp(-2kt). \eqno{(13)}
$$
From Eq.(13), we can easily see that the entanglement of
$|B^1\rangle$ and $|B^2\rangle$ are more fragile than entanglement
of $|B^3\rangle$ and $|B^4\rangle$ in the independent dissipative
environment. In the above discussion, we have set $\gamma=0$,
i.e., not any direct coupling between cavity $A$ and $B$. If
$\gamma\neq0$, the time evolution density matrices corresponding
initial $|B^{1}\rangle$ or $|B^{2}\rangle$ can not be regarded as
 mixed two-qubit states but mixed two-qutrit states [20]. However,
the time evolution density matrices corresponding initial
$|B^{3}\rangle$ or $|B^{4}\rangle$ are still two-qubit mixed
states, and their concurrences given by $\exp(-2kt)$ are
independent with the coupling constant $\gamma$.\\
\hspace*{8mm}For another initial condition of interest, i.e., the
cavity $A$ is prepared in a mesoscopic superposition state
$N(\theta,\phi)(|\alpha{e}^{i\phi}\rangle+e^{i\theta}|\alpha{e}^{-i\phi}\rangle)$,
where
$$
N(\theta,\phi)=[2+2\cos(\theta-|\alpha|^2\sin2\phi)\exp(|\alpha|^2\cos2\phi-|\alpha|^2)]^{-\frac{1}{2}}
\eqno{(14)}
$$
is the normalization constant and $|\alpha{e}^{\pm{i}\phi}\rangle$
is the coherent state. The cavity $B$ is in the vacuum state. The
time evolution density matrix $\rho(t)$ can be expressed as,
$$
\rho(t)=N^2(\theta,\phi)\{|\alpha_{+}(t)\rangle\langle\alpha_{+}(t)
|\otimes|\beta_{+}(t)\rangle\langle\beta_{+}(t)|+
|\alpha_{-}(t)\rangle\langle\alpha_{-}(t)|\otimes|\beta_{-}(t)\rangle\langle\beta_{-}(t)|
$$
$$
~~~+[\xi(t)|\alpha_{+}(t)\rangle\langle\alpha_{-}(t)|
\otimes|\beta_{+}(t)\rangle\langle\beta_{-}(t)|+h.c.]\},
\eqno{(15)}
$$
where
$$
\alpha_{\pm}(t)=\alpha{e}^{\pm{i}\phi-i\omega{t}-kt}\cos\gamma{t},
~~~\beta_{\pm}(t)=-i\alpha{e}^{\pm{i}\phi-i\omega{t}-kt}\sin\gamma{t},
$$
$$
\xi(t)=e^{-i\theta}\exp[(1-e^{-2kt})(e^{2i\phi}-1)|\alpha|^2].
\eqno{(16)}
$$
Since the density operator in Eq.(15) can be considered as a mixed
two-qubit state [21], we adopt the entanglement of formation [22]
to calculate the entanglement between cavities $A$ and $B$. For a
mixed two-qubit state, the entanglement of formation $E$ related
to the concurrence $C$ by
$E=h(\frac{1}{2}+\frac{1}{2}\sqrt{1-C^2})$, where $h$ is the
binary entropy function $h(x)=-x\log_2x-(1-x)\log_2(1-x)$ [19].\\
\begin{figure}
\centering
\includegraphics{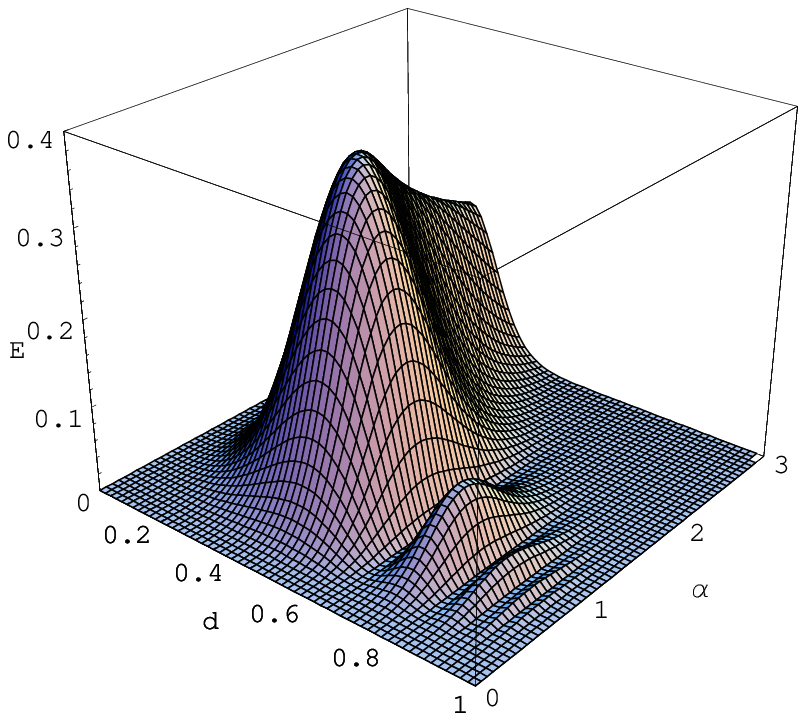}
\caption{The entanglement of formation $E$ of the mixed state
$\rho(t)$ is plotted as a function of the degree of decay $d$ and
the initial amplitude $\alpha$ of the mesoscopic superposition
state for $\gamma/k=6$, $\phi=\pi/2$ and $\theta=0$.
\label{Fig.1}}
\end{figure}
\hspace*{8mm}In Fig.1, the entanglement of formation $E$ of the
mixed state $\rho(t)$ is plotted as a function of the degree of
decay $d=\sqrt{1-e^{-2kt}}$ and the initial amplitude $\alpha$ of
the mesoscopic superposition state. It is shown that the
entanglement between cavity $A$ (which is initially in even
coherent state) and cavity $B$ (initially in vacuum state) arises
in the dissipative process. The maximal value of entanglement
achieved during the evolution firstly increases with the initial
amplitude of the even coherent state, then decay slowly. In some
appropriate values of the amplitude, the entanglement exhibit the
revival and collapse phenomenon. If the initial amplitude of the
even coherent state is large enough, the revival of entanglement
does not appear.\\
\begin{figure}
\centering
\includegraphics{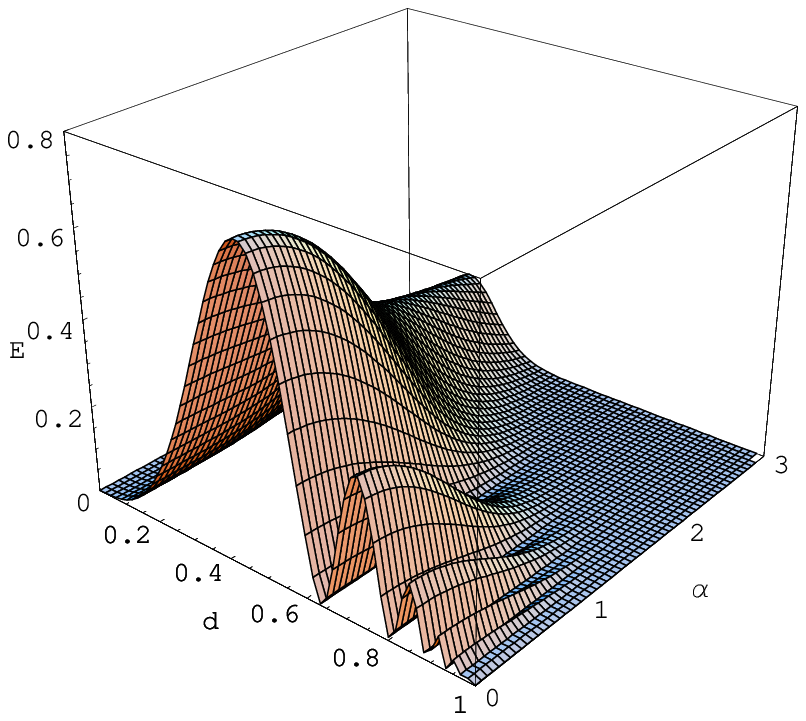}
\caption{The entanglement of formation $E$ of the mixed state
$\rho(t)$ is plotted as a function of the degree of decay $d$ and
the initial amplitude $\alpha$ of the mesoscopic superposition
state for $\gamma/k=6$, $\phi=\pi/2$ and $\theta=\pi$.
\label{Fig.2}}
\end{figure}
\hspace*{8mm}The entanglement of formation $E$ of the mixed state
$\rho(t)$ is plotted as a function of the degree of decay
$d=\sqrt{1-e^{-2kt}}$ and the initial amplitude $\alpha$ in Fig.2.
In this case, the initial superposition state in cavity $A$ is a
odd coherent state. We can see that a different feature of
entanglement emerges in this case compared with the one in Fig.1.
The entanglement decreases with the increase of the amplitude of
the odd coherent state. For very small initial amplitude, the
entanglement exhibits the revival and collapse due to the coupling
between the cavity $A$ and cavity $B$.
\begin{figure}
\centering
\includegraphics{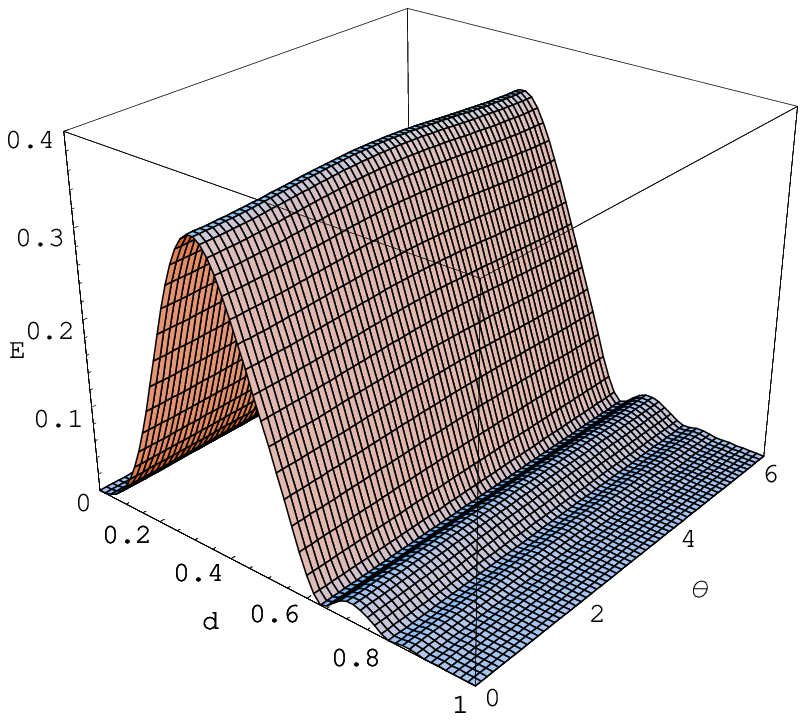}
\caption{The entanglement of formation $E$ of the mixed state
$\rho(t)$ is plotted as a function of the degree of decay $d$ and
the initial relative phase factor $\theta$ of the mesoscopic
superposition state for $\gamma/k=6$, $\phi=\pi/2$ and
$\alpha=1.5$. \label{Fig.3}}
\end{figure}
In Fig.3, the entanglement of formation $E$ of the mixed state
$\rho(t)$ is plotted as a function of the degree of decay
$d=\sqrt{1-e^{-2kt}}$ and the initial relative phase factor
$\theta$ of the mesoscopic superposition state. It is shown that,
for large value of the initial amplitude of superposition state,
the entanglement is not heavily dependent of the relative phase
$\theta$ of the mesoscopic superposition state. However, for very
small value of the initial amplitude, the entanglement generation
heavily depends on the relative phase $\theta$, which can be
observed in Fig.4.\\
\begin{figure}
\centering
\includegraphics{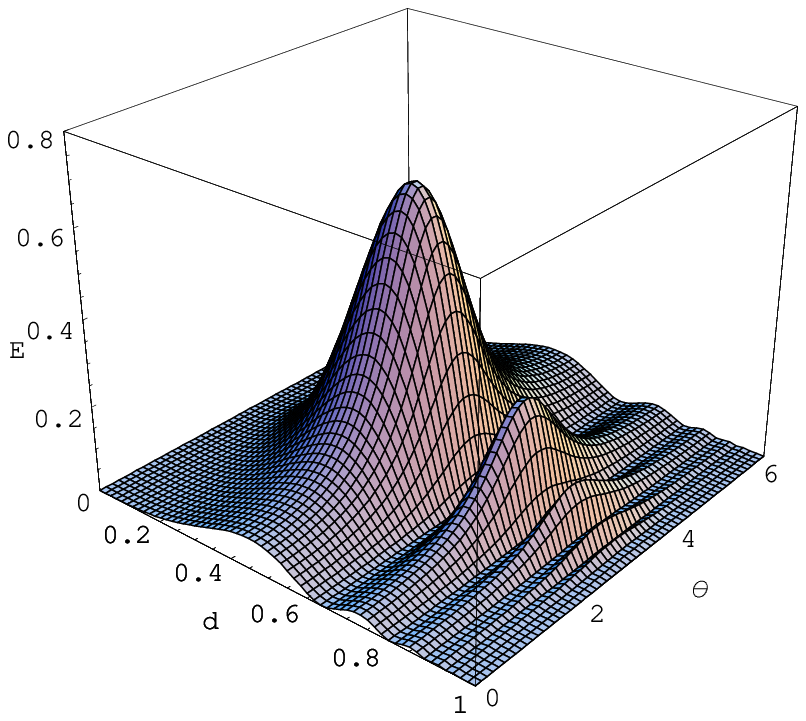}
\caption{The entanglement of formation $E$ of the mixed state
$\rho(t)$ is plotted as a function of the degree of decay $d$ and
the initial relative phase factor $\theta$ of the mesoscopic
superposition state with $\gamma/k=6$, $\phi=\pi/2$, $\alpha=0.5$.
\label{Fig.4}}
\end{figure}
\hspace*{8mm}Recently, much attention has been paid to the
theoretical studies on quantum nonlocality for continuous
variable. Banaszek and W{\' o}dkiewicz have developed a Wigner
function representation of the Bell-Clauser, Horne, Shimony and
Holt (CHSH) [23] inequality using a two-mode parity operator
$\Pi(\mu,\nu)$ as a quantum observable [14,15]. In what follows,
we investigate the dynamic behavior of the quantum nonlocality
based on parity measurement for the two coupled cavity fields in a
dissipative environment. The type of quantum observable to be
measured is a crucial factor in the nonlocality test. The two-mode
displaced parity operator $\Pi(\mu,\nu)$ is defined by
$$
\Pi(\mu,\nu)=D_1(\mu)D_2(\nu)\Pi D_1^\dagger(\mu)D_2^\dagger(\nu),
\eqno{(17)}
$$
where $D_1(\mu)$ and $D_2(\nu)$ are the displacement operators of
the mode 1 and the mode 2, respectively, and
$$
\Pi= \Pi_{e1}\otimes\Pi_{e2}-\Pi_{e1}\otimes\Pi_{o2}-\Pi_{o1}\otimes\Pi_{e2}+\Pi_{o1}\otimes\Pi_{o2},\label{eq:pm2}\\
\eqno{(18)}
$$
with
$$
\Pi_e=\sum_n^\infty|2n\rangle\langle 2n|,~~
\Pi_o=\sum_n^\infty|2n+1\rangle\langle 2n+1|. \eqno{(19)}
$$
The Bell-CHSH inequality is then
$$
|{\cal
B}|=|\langle\Pi(\mu,\nu)+\Pi(\mu,\nu^\prime)+\Pi(\mu^\prime,\nu)-\Pi(\mu^\prime,\nu^\prime)\rangle|\leq2,
\eqno{(20)}
$$
where we call $|{\cal B}|$ the Bell measure. The displacement
operation can be effectively performed using a beam splitter with
the transmission coefficient close to one and a strong coherent
state being injected into the other input port [15]. The two-mode
Wigner function at a given phase point described by $\mu$ and
$\nu$ is [24,16]
$$
W(\mu,\nu)=\frac{4}{\pi^2}{\rm Tr}[\rho\Pi(\mu,\nu)], \eqno{(21)}
$$
where $\rho$ is the density operator of the two cavity fields.
Eqs.(20) and (21) lead to the Wigner representation of Bell's
inequality
$$
|{\cal
B}|=\frac{\pi^2}{4}|W(\mu,\nu)+W(\mu,\nu^\prime)+W(\mu^\prime,\nu)
-W(\mu^\prime,\nu^\prime)|\leq2. \eqno{(22)}
$$
Measurement of the degree of quantum nonlocality is defined by the
maximal violation of Bell's inequality (22). The Wigner function
of the density operator $\rho(t)$ in Eq.(15) can be derived from
the Fourier transform of it's characteristic function
$$
C(\eta,\zeta)={\rm Tr}[\rho D_1(\eta)D_2(\zeta)]. \eqno{(23)}
$$
Then, we obtain the Wigner function of the density operator
$\rho(t)$ as follows,
$$
W(\mu,\nu)=N^2(\theta,\phi)\{\frac{4}{\pi^2}\exp(-2|\mu-\alpha_{+}(t)|^2
-2|\nu-\beta_{+}(t)|^2)+\frac{4}{\pi^2}\exp(-2|\mu-\alpha_{-}(t)|^2
-2|\nu-\beta_{-}(t)|^2)
$$
$$
~~~+[\frac{4}{\pi^2}\xi(t)\exp[i{\mathrm{Im}}(\mu\alpha^{\ast}_{+}(t)
-\mu\alpha^{\ast}_{-}(t))-\frac{1}{2}|\mu+\alpha_{+}(t)|^2-\frac{1}{2}|\mu
+\alpha_{-}(t)|^2-(\mu+\alpha_{+}(t))(\mu^{\ast}+\alpha^{\ast}_{-}(t))]
$$
$$
\cdot\exp[i{\mathrm{Im}}(\nu\beta^{\ast}_{+}(t)
-\nu\beta^{\ast}_{-}(t))-\frac{1}{2}|\nu+\beta_{+}(t)|^2-\frac{1}{2}|\nu+\beta_{-}(t)|^2
-(\nu+\beta_{+}(t))(\nu^{\ast}+\beta^{\ast}_{-}(t))]+c.c.]\}.
\eqno{(24)}
$$
In order to see the evolution of the nonlocality, we calculate the
the maximum of the Bell measure by the steepest descent method
[25]. The result (According to Banaszek and W{\'o}dkiewicz [15],
we choice the parameters $\mu=\nu=0$ in Eq.(22)) is displayed in
Fig.5. It is found that, the two cavity fields become nonlocal in
the beginning of the evolution, then, their nonlocality disappear
as time proceeds. The maximal value of $|{\cal B}|_{max}$ achieved
by two fields depends on three parameters, i.e., the initial
amplitude $|\alpha|$, the relative phase $\theta$, and the rate
$\gamma/k$. Roughly speaking, it increases with $|\alpha|$ and
$\gamma/k$. If the initial amplitude is small enough and the rate
$\gamma/k$ is large enough, the revival of the Bell violation can
be observed. We have also calculated the maximal violation in the
case with $\theta=0$ and $|\alpha|=0.5$, and found that the
violation is more robust against the dissipation than the one in
the case with $\theta=\pi$
and $|\alpha|=0.5$, although its maximal value of violation is lowered.\\

In conclusion, we investigate the entanglement between two
distinct coupled cavities, in which one is initially in the
schr\"{o}dinger cat state and the other is in the vacuum in the
dissipative environment. We find that the relative phase of the
cat state play a very sensitive role in the entanglement if the
initial amplitude of the schr\"{o}dinger cat state is very small.
It is also shown that the entanglement between two cavities (in
which one is initially in even coherent state and the other in
vacuum state) arises in the dissipative process. The maximal value
of entanglement achieved during the evolution firstly increases
with the initial amplitude of the even coherent state, then decay
slowly. In some appropriate values of the amplitude, the
entanglement exhibit the revival and collapse phenomenon. The
entanglement decay of the four Bell states in the cavities
surrounded by dissipative environment is also discussed and two
kinds of entanglement decay rates are found. Finally, the dynamic
behavior of nonlocality for the two coupled cavity in dissipative
environment is examined. We show that the two cavity fields can
violate the Bell-CHSH inequality during the evolution in the
dissipative environment, then, the Bell violation disappears as
time proceeds. The results obtained in this Letter may have some
applications to the quantum computation based on the coherent state [26,27].\\

\begin{figure}
\centering
\includegraphics{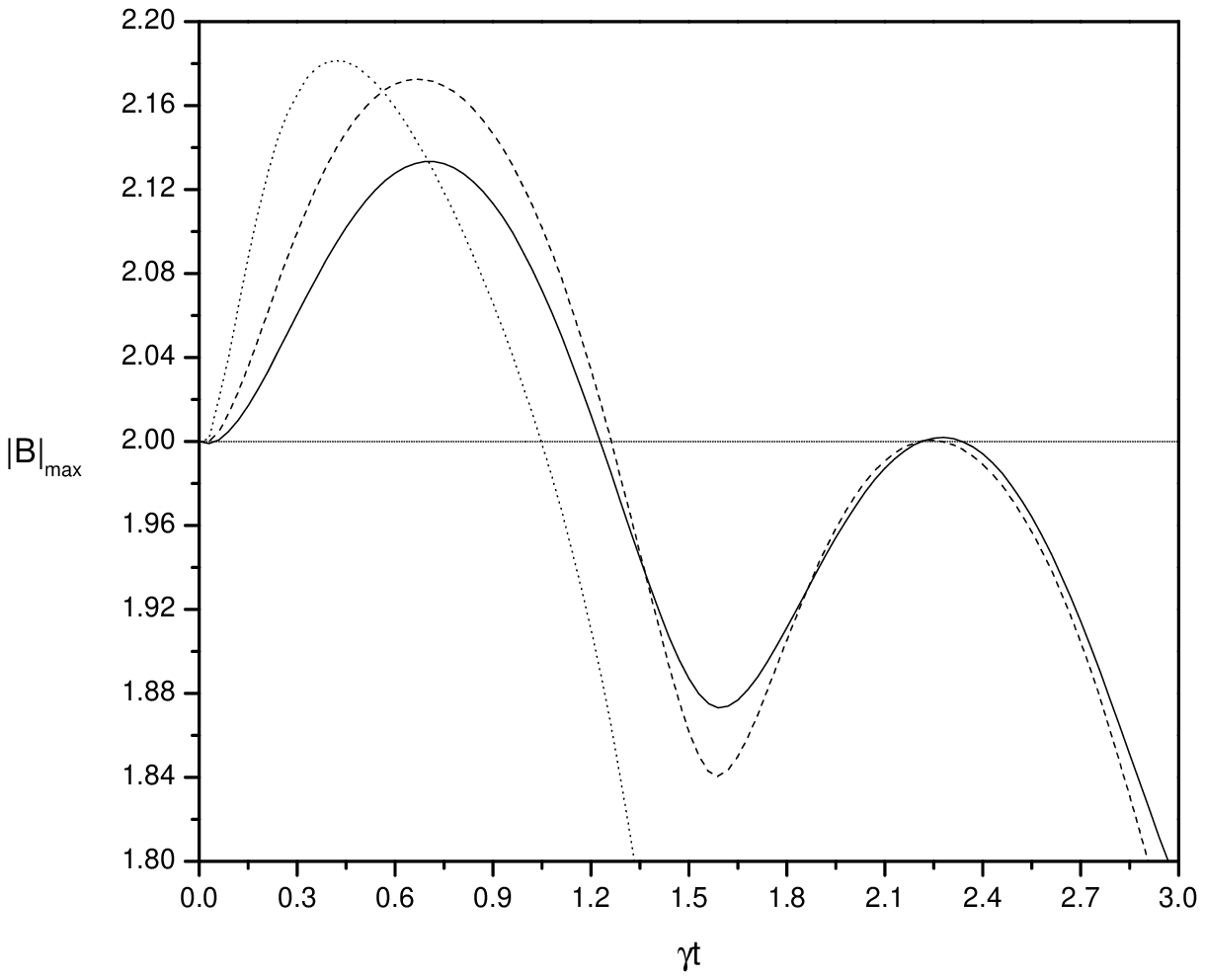}
\caption{The time evolution of $|B|_{max}$ is plotted as the
function of $\gamma{t}$ with $\gamma/k=100$, $\phi=\pi/2$,
$\theta=\pi$ for three different initial amplitudes:
$|\alpha|=0.5$ (Solid Line); $|\alpha|=1$ (Dash Line);
$|\alpha|=2$ (Dot Line). \label{Fig.5}}
\end{figure}

\section * {ACNOWLEDGMENT}
This project was supported by the National Natural Science
Foundation of China (Project NO. 10174066).

\end{document}